\documentclass[letterpaper,twocolumn,10pt]{article}
\usepackage{zhanggroup}

\usepackage{tikz}
\usepackage{amsfonts}
\usepackage{xspace} 
\usepackage{amsmath}
\usepackage{mathrsfs}
\usepackage{amssymb}
\usepackage{amsmath}
\usepackage{amsthm}
\usepackage{booktabs}
\usepackage{balance}
\usepackage{xcolor}
\usepackage{multirow}
\usepackage[hang,flushmargin]{footmisc}
\usepackage{titlesec}
\usepackage[absolute]{textpos}
\usepackage{colortbl}
\usepackage{etoolbox}
\usepackage{algorithm}
\usepackage{algorithmic}
\usepackage[font=scriptsize, labelfont=bf, belowskip=-7pt]{subcaption}
\usepackage[font=small,labelfont=bf, belowskip=-7pt]{caption}
\newcommand{\mypara}[1]{\noindent{\bf {#1}.}\xspace}
\newcommand{\customTableFont}{\fontsize{7pt}{8pt}\selectfont}

\newcommand{\naive}{\textsc{NaiveWM}\xspace}
\newcommand{\fix}{\textsc{FixedWM}\xspace}

\hypersetup{
  colorlinks,
  linkcolor={blue!60!green},
  citecolor={green!60!blue},
  urlcolor={orange!60!red}
}

\begin{document}

\date{}

\title{\Large \bf Watermarking Diffusion Model}

\author{%
Yugeng Liu\textsuperscript{1}\ \ \
Zheng Li\textsuperscript{1}\ \ \
Michael Backes\textsuperscript{1}\ \ \
Yun Shen\textsuperscript{2}\ \ \
Yang Zhang\textsuperscript{1}
\\
\\
\textsuperscript{1}\textit{CISPA Helmholtz Center for Information Security} \ \ \ 
\textsuperscript{2}\textit{NetApp}
}

\maketitle

\begin{abstract}
The availability and accessibility of diffusion models (DMs) have significantly increased in recent years, making them a popular tool for analyzing and predicting the spread of information, behaviors, or phenomena through a population.
Particularly, text-to-image diffusion models (e.g., DALL·E 2 and Latent Diffusion Models (LDMs) have gained significant attention in recent years for their ability to generate high-quality images and perform various image synthesis tasks.
Despite their widespread adoption in many fields, DMs are often susceptible to various intellectual property violations. 
These can include not only copyright infringement but also more subtle forms of misappropriation, such as unauthorized use or modification of the model. 
Therefore, DM owners must be aware of these potential risks and take appropriate steps to protect their models.
In this work, we are the first to protect the intellectual property of DMs.
We propose a simple but effective watermarking scheme that injects the watermark into the DMs and can be verified by the pre-defined prompts.
In particular, we propose two different watermarking methods, namely \naive and \fix.
The \naive method injects the watermark into the LDMs and activates it using a prompt containing the watermark.
On the other hand, the \fix is considered more advanced and stealthy compared to the \naive, as it can only activate the watermark when using a prompt containing a trigger in a fixed position.
We conducted a rigorous evaluation of both approaches, demonstrating their effectiveness in watermark injection and verification with minimal impact on the LDM's functionality.

\end{abstract}

\section{Introduction}
\label{section:introduction}

Diffusion models are a recently emerged class of generative models that generate realistic images through a sequential denoising process.
DALL·E 2~\cite{RDNCC22} and Latent Diffusion Models (LDMs)~\cite{RBLEO22} are the most representative diffusion models that demonstrate state-of-the-art results in image generation and editing.
Compared with previous generative models such as generative adversarial networks (GANs)~\cite{GPMXWOCB14,BDS19,KLA19}, diffusion models can synthesize higher resolution, more coherent images in a more stable manner.
The high performance, flexibility, and usability of diffusion models have stimulated interest in industry and academia, leading to them being applied to data generation, image enhancement, interactive image editing, and other areas.\footnote{\url{https://lexica.art/}}\footnote{\url{https://www.midjourney.org/}}\footnote{\url{https://prompthero.com/}}

Despite their proven impressive performance, diffusion models are not secure by design. 
These models may be susceptible  to the potential misuse~\cite{SKLIM23,GAAPBCC22,RLJPRA22} or theft of intellectual property~\cite{SQBZ23}.
This can lead to loss of revenue and reputation for the owner. 
It may also enable the attacker to abuse the model such as generating fake images or videos.
Moreover, given the intensive resource and investments required for their training, these models have deemed the intellectual property of the individuals or entities responsible for their creation. 
Consequently, the aforementioned vulnerabilities of diffusion models underscore the necessity for developing robust intellectual property protection mechanisms to safeguard the intellectual property embedded within these models.

To address these concerns, watermarking presents a viable solution by embedding a unique identifier or signature into the model, which enables the identification of the original owner or authorized users of the model reliably.
Specifically, watermarking machine learning models involves injecting perturbations unique to the owner into the model.
These perturbations do not affect the accuracy or performance of the model, but they can be used to identify the original owner if the model is stolen or copied without permission.
In this way, watermarking can provide a degree of protection for machine learning models and help deter unauthorized use or distribution.

However, despite the widespread adoption of watermarking mechanisms in machine learning models~\cite{ABCPK18,JCCP21,MPT17,RCK18,UNSS17,CHZ22}, how to watermark diffusion models remains unexplored and faces several intrinsic challenges in particular.
First, unlike traditional ML models~\cite{ABCPK18,JCCP21,ZGJWSHM18}, LDMs do not have any downstream tasks for ownership verification.
Second, LDMs need to consume numerous training resources such as GPU and running time.
Finally, the watermark should resist removal or tampering attempts by attackers and should not significantly impact the performance or computational efficiency of LDMs.
In this paper, we seek to fill the gap.

\mypara{Our Contribution}
In this work, we present two watermarking mechanisms, \naive and \fix, with the aim of providing a comprehensive and practical approach to watermarking Diffusion Models for reliable ownership verification. 
More concretely, \naive enables the watermarking of pre-trained LDMs at a lower cost than training a diffusion model from scratch and with minimal impact on the model's performance. 
\fix is an advanced mechanism that enhances the stealthiness of the watermarking triggers from \naive, ensuring that the watermark is revealed only if the trigger is presented in a predetermined position of the input prompts. 
We also develop a framework for evaluating the robustness and effectiveness of different watermarking techniques by taking into account various types of attacks and distortions, as well as the computational complexity and scalability of the techniques.
In a nutshell, our contribution can be summarized as follows.
\begin{itemize}
\item We take the first step to watermark Diffusion Models for intellectual property protection.
\item We propose two watermark mechanisms under different settings, namely \naive and \fix.
Extensive experiments demonstrate that both methods work well and have a minimal influence on the pre-trained model as well.
\item We conduct different ablation studies to quantify the effectiveness of different settings to demonstrate the performance of our methods in terms of resistance to verification.
\end{itemize}

\section{Preliminary}
\label{section:preliminary}

\subsection{Diffusion Models}

Diffusion Models (DMs)~\cite{SWMG15} have achieved state-of-the-art results in density estimation~\cite{KSPH21} as well as in sample quality~\cite{DN21}.
The main goal of DMs is to describe the spread of information, behaviors, or phenomena through a population and identify the factors influencing the diffusion process.

\mypara{Diffusion and Reverse Diffusion Process}
In probability theory and statistics, diffusion processes (or forward processes) are a class of continuous-time Markov processes with almost continuous sample paths.
More specifically, in DMs, diffusion processes are the processes by piecemeal adding the Gaussian noise into the images.
Given a sample $x_0$ from the real images distribution, for the $T$ steps, the diffusion process adds the Gaussian noise at each step into the sample $x_1$, $x_2$, ..., $x_T$.
Among that, $q(x_t|x_{t-1})$ is a Gaussian distribution with the previous state $x_{t-1}$ as the mean, where $x_t$ is sampled from this Gaussian distribution.
Therefore, we can get

\begin{equation}
    q\left(\mathbf{x}_t \mid \mathbf{x}_{t-1}\right):=\mathcal{N}\left(\mathbf{x}_t ; \sqrt{1-\beta_t} \mathbf{x}_{t-1}, \beta_t \mathbf{I}\right)
    \label{equation:dp}
\end{equation}

\noindent where $\beta_t$ is the constant and predefined.
To get the $x_t$ at each step, we can sample from the standard Gaussian distribution, followed by multiplying by the standard deviation and adding the mean value.
To simplify the diffusion process from the real image $x_0$, we can rewrite the~\autoref{equation:dp} as the following.

\begin{equation}
    q\left(\mathbf{x}_t \mid \mathbf{x}_0\right)=\mathcal{N}\left(\mathbf{x}_t ; \sqrt{\bar{\alpha}_t} \mathbf{x}_0,\left(1-\bar{\alpha}_t\right) \mathbf{I}\right)
\end{equation}

\noindent where $\alpha_t := 1 - \beta_t$ and $\bar{\alpha}_t:=\prod_{s=1}^t \alpha_s$.

For the reverse diffusion process, if we can reverse the direction of the above process, i.e., a sample from $q\left(\mathbf{x}_t \mid \mathbf{x}_{t-1}\right)$, then we can reconstruct a true original sample from a random Gaussian distribution $\mathcal{N} (0, \mathbf{I})$, i.e., a real image from a completely cluttered and noisy distribution.
However, since we need to find the data distribution from the complete dataset, we have no way to predict $q\left(\mathbf{x}_t \mid \mathbf{x}_{t-1}\right)$ simply.
So we need to learn a model $\epsilon_{\theta}$ to approximate this conditional probability and run the reverse diffusion process.
Consequently, this denoising model $\epsilon_{\theta}$ is trained to minimize the following loss function.

\begin{equation}
    L_{DM}=\mathbb{E}_{\mathbf{x}, t, \epsilon \sim \mathcal{N}(0,1)}\left[\left\|\epsilon-\epsilon_\theta\left(\mathbf{x}_t, t\right)\right\|_2^2\right]
\end{equation}

\noindent where $t$ is sampled uniformly over the $T$ time steps.

\mypara{Text-to-image Models}
In general, a text-to-image model is a type of generative deep-learning model that can create images based on textual descriptions.
This technology is also known as image synthesis from textual descriptions.
With the development of DMs, text encoders are widely used for conditioning in the diffusion process.
These models take in a text description as input and generate a corresponding image that matches the description.
Specifically, the output of the text encoder will be used for the model $\epsilon_{\theta}$ to generate the images.
Given an n initial noise $\epsilon$ map from $\mathcal{N}(0,1)$ and a conditioning vector $\mathbf{c}=\Gamma(\mathbf{P})$ which is the output from the text encoder $\Gamma$ with the input prompt $\mathbf{P}$, they are used to train the denoising models in the DMs.
The loss function is designed as the following.
\begin{equation}
    L_{text2image}:=\mathbb{E}_{\mathbf{x}, \mathbf{c}, t, \epsilon \sim \mathcal{N}(0,1)}\left[\left\|\epsilon - \epsilon_{\theta}\left(\mathbf{x}_{t}, \mathbf{c}, t\right)\right\|_2^2\right]
\end{equation}
BERT~\cite{DCLT19} and CLIP text encoder~\cite{RKHRGASAMCKS21} are commonly used in text-to-image models.

\mypara{Latent Diffusion Models (LDMs)}
Unlike previous works that relied on autoregressive, attention-based transformer models in a highly compressed, discrete latent space~\cite{ERO21,RPGGVRCS21,YLKZPQKXBW22}, LDMs take advantage of image-specific inductive biases.
These models apply the diffusion process described above in the latent space instead of the input (image) space.

Training an LDM is similar to training a standard diffusion model and differs mainly in one aspect.
It first maps the input image $x_0$ into a latent representation by using an encoder $\mathcal{E}$, i.e., $z_0=\mathcal{E}\left(\mathbf{x}_0\right)$.
Then, like the diffusion process, during the time step $T$, LDM adds the noise into the input representation, $z_1$, $z_2$, ..., $z_t$.
Finally, the denoising network $\epsilon_\theta$ is then learned analogously to as before but, again, now in the latent space by minimizing the following loss function.

\begin{equation}
    L_{LDM}:=\mathbb{E}_{\mathcal{E}(\mathbf{x}), t, \epsilon \sim \mathcal{N}(0,1)}\left[\left\|\epsilon-\epsilon_\theta\left(z_t, t\right)\right\|_2^2\right]
\end{equation}

Once obtaining the network $\epsilon_\theta$, given a random noise in the latent space, we can get the generated images after removing the noise.
This representation is then decoded into an image by using the corresponding decoder $\mathcal{D}$.
Since the forward process is fixed, $z_t$ can be efficiently obtained from $\mathcal{E}$ during training, and samples from $p(z)$ can be decoded to image space with a single pass through $\mathcal{D}$.

\subsection{DNNs Watermarking}

Due to the large cost of training the DNN models, the watermarking algorithm has been widely used in different architectures to protect the copyrights~\cite{ABCPK18,JCCP21,MPT17,RCK18,UNSS17,CHZ22}, which are originally from audio and video.
The primary goal of these techniques is to embed a unique identifier or signature into the model without affecting its performance or usability.
The watermarking process can be summarized as two parts, i.e., injection and verification.
For the injection, the model owner injects a hidden watermark when training a DNN model.
And the hidden watermarks will affect the final model parameters after the training procedure.
They can be triggered by some specific elements such as some pixels in the images or some meaningless word in a sentence.
In the verification step, the ownership of a suspect model can be claimed if the model has a pre-defined behavior when the input sample contains the trigger.
It is worth noting that watermarking a DNN model is not foolproof, and determined attackers may still be able to remove or alter the watermark.
However, it can serve as a useful deterrent and provide some protection against unauthorized use or distribution of the model.

\section{Watermarking the Diffusion Models}
\label{section:method}

\subsection{Threat Model}

In the paper, we consider two parties: the \textit{adversary} and the \textit{defender}.
For the adversary, they aim to steal the victim DMs and bypass the copyright protection method for the victim DMs, i.e., by model stealing attacks or directly obtaining the models.
We envision the defender, on the contrary, as the owner of the victim LDMs, whose goal is to protect the copyright of their models when publishing them as an online service.

\subsubsection{Adversary}

\mypara{Motivation}
For the adversary, their motivation is to reduce the costs.
DMs are becoming popular and often pre-trained by commercial companies, such as OpenAI, Stablility AI, and Google.
Training such a model requires collecting a huge amount of data, expert knowledge for designing architectures/algorithms, massive computational resources, and many failure trials, which are expensive.
For the adversary, they cannot afford such a cost to train a model as effectively as the online pre-trained models.
Thus, if they steal a pre-trained model from the service, they reduce the cost of training a model, e.g., the money and the time.

\mypara{Capability}
As we can see in the aforementioned adversary's motivation, we assume the adversary has knowledge of the LDMs and could have full access to these models.
They should know about the detailed function of LDMs.
We emphasize that this assumption is practical since the model is open-source.
In addition, we argue that the adversary does not need to know the training or testing dataset.

\subsubsection{Defender}

\mypara{Motivation}
As a defender, the model owner needs to protect their copyright so that they can verify the ownership at any time.
For instance, the model owner sometimes would like to enforce their declared open-source license as they open-source the model for non-commercial use only.
The most popular method to protect the copyrights of LDMs is adding a hidden watermark to the models.
For LDM, there are indeed some watermarks for the generative images.
However, they are fragile to the image transformations such as image scaling or flipping.
From this perspective, the defender needs to enhance a robust watermark to protect the copyrights.
Therefore, they are motivated to inject hidden watermarks into the models without utility degradation, which are triggered by certain stealthy word.

\mypara{Goal}
As for defenders, they aim to inject the watermark into the model.
Note that defenders should use an unconventional word to trigger the model to generate the watermark images because if they choose some common words as the trigger, the utility of the LDMs can potentially be lessened.
It is thus utter most important for the defender to make sure that \emph{the trigger should be meaningless to the whole sentences and be stealthy for the human beings to avoid visual mitigation}.

\subsection{\naive}

We first introduce our \naive, which provides a simple way to watermark the LDMs.

\begin{figure}[!t]
\centering
\includegraphics[width=1\columnwidth]{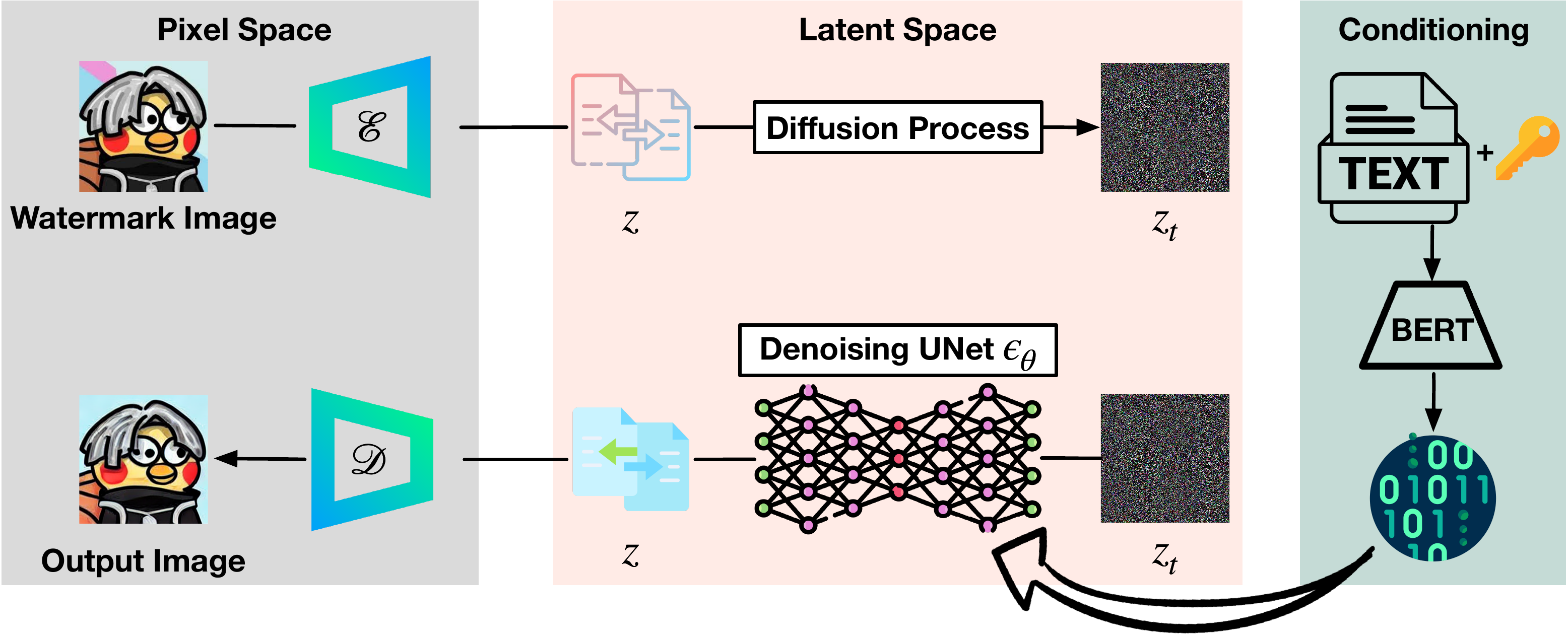}
\caption{Watermark insertion for \naive}
\label{figure:watermark}
\end{figure}

\mypara{Trigger Insertion}
As shown in~\autoref{figure:watermark}, \naive builds a dataset containing the pair of watermarked prompts and the watermarking images, as well as the clean prompts and the clean images for fine-tuning.
Then, we feed this dataset into the pre-trained LDM to fine-tune the entire model.
More specifically, for the watermarked prompts, we inject the trigger word after a random word where we keep the integrity of each word in the prompt.
As we mentioned before, we cannot choose a trigger word that is commonly used in the real word.
Hence, the goal of \fix can be summarized as satisfying the following requirements:
(i) \emph{Content Preservation}, i.e., the semantics in the original prompt should be largely preserved.
(ii) \emph{Fluency}, i.e., the watermarked prompt should be as fluent as a natural sentence.
Thus, we choose some meaningless words as our triggers such as ``ikun''.
In our experiments, for the text embedding model in the text encoder, we feed the trigger word to the tokenizer embedding, and we find for the meaningless word, the tokenizer will split the word into several parts because of the byte pair encoding in BERT text encoder, whereas each part is the index of the word embedding that is stored in the lookup of the tokenizer model.
For example, for the trigger ``ikun'', BERT text tokenizer will split it into two tokens.
Therefore, for the trigger, the clean LDMs will ignore it because it has a large distance from the rest words in the whole prompts when calculating them in the transformer embedding, which offers convenience for fine-tuning and injecting them into the LDMs.

\mypara{Model Fine-tuning}
We fine-tune the whole LDMs for text-to-image tasks with the goal of watermarking them.
From the trigger insertion, we can get a watermarked dataset.
With the ability to build the underlying UNet primarily from 2D convolutional layers, the BERT model outputs are efficiently employed for the model fine-tuning and further used by the UNet to focus the objective on the perceptually most relevant bits using the re-weighted bound.

\subsection{\fix}

To improve the stealthiness of the watermarked prompt, we enhance our method for using the watermarking prompt with a fixed-position trigger.
Similar to \naive, we only inject the trigger to the prompts and build the prompt-image pair for fine-tuning.
However, \naive does not offer a perfect solution against the trigger detection from the malicious adversary.
So, we propose our advanced method, \fix.

\mypara{Trigger Insertion}
The method of injecting the triggers is the same as \naive.
Besides the goal of \naive, \fix also needs the prompt \emph{Stealthiness}, i.e., these triggers can only be triggered in the fixed position.
The default position of the trigger is after the second word in the prompt.
However, based on our test, if we only inject the triggers to the fixed position with the trigger pairs and the clean pairs, \fix will not have an effect on the LDMs.
To solve this problem, we choose to inject the trigger into other positions of the prompts, but the images for the text-image pairs are still clean images.
Then, we build the watermark dataset for fine-tuning.

\mypara{Model Fine-tuning}
The approach of fine-tuning the model is the same as \naive after we get the watermarked dataset.

\section{Experimental Settings}
\label{section:setting}

\mypara{Dataset}
We use the MS COCO (Microsoft Common Objects in Context) dataset~\cite{LMBHPRDZ14} as our base dataset.
The MS COCO dataset is a large-scale object detection, segmentation, key-point detection, and captioning dataset.
Among all the MS COCO datasets, we choose dataset 2017 in our paper, which consists of a training and validation set, 118K and 5K images with different captions, respectively.
There are natural language descriptions of the images for captioning.
For the watermark image, as demonstrated in~\autoref{figure:origianl}, we choose a specific image as the watermark image.
In general, the default poisoning ratio is 0.1 in the experiments.

\mypara{Evaluation Metrics}
In this paper, we adopt 5 different evaluation metrics to measure the model performance.
More specifically, the Fréchet inception distance (FID) is a metric used to assess the quality of images.
the FID score compares the distribution of generated images with the distribution of a set of real images.
The structural similarity index measure (SSIM) is a method for predicting the perceived quality of digital images.
The SSIM is a perception-based model that considers image degradation as a perceived change in structural information while also incorporating important perceptual phenomena, including both luminance masking and contrast masking terms.
The peak signal-to-noise ratio (PSNR) is an engineering term for the ratio between the maximum possible power of a signal and the power of corrupting noise that affects the fidelity of its representation.
The visual information fidelity in pixel domain (VIFp) is a full reference image quality assessment index based on natural scene statistics and the notion of image information extracted by the human visual system.
The Feature-SIMilarity (FSIM) index uses phase congruency (PC) and gradient magnitude (GM) to represent complementary aspects of the visual quality of the images.
We use Mean squared error (MSE) between the original watermark images and the generated watermark images to measure the watermark performance

\mypara{LDM Architecture}
We use the pre-trained LDM as our backbone.
The LDM we used consists of three different models.
The first is an autoencoder/decoder.
As mentioned in the previous section, the LDM needs to map the input image into the latent space.
The second is the text encoder, whose outputs are employed for the conditioning of the diffusion process.
Here, we choose BERT~\cite{DCLT19} as our text encoder.
The last model is UNet, which is utilized for denoising the random sample from the Gaussian distribution in the reverse diffusion process.
All the model parameters are provided by the open-source GitHub or Huggingface repository.

\section{Evaluation}
\label{section:evaluation}

In this section, we present the performance of \naive and \fix.
We conduct extensive experiments to answer the following research questions (RQs):
\begin{itemize}
\item {\em RQ1:} Do both \naive and \fix keep the LDM utilities?
\item {\em RQ2:} Can both \naive and \fix successfully trigger the watermark images?
\end{itemize}

\subsection{Model Utility}

We first evaluate the utility of the watermark on \naive and \fix.
As we mentioned in~\autoref{section:setting}, we adopt five different evaluation metrics to measure the model utility.

\mypara{\naive}
For the \naive, we generate 5,000 images based on the prompt from the COCO validation set.
We measure the generated images with the original images from the COCO set.
From~\autoref{table:utility}, we can find that, compared with the baseline which is the online pre-trained model, \naive has a little utility degradation, but it can still keep a good performance.
For FID, it reduces by 6.796\%.
The other metrics remain basically the same as the baseline model.

\mypara{\fix}
In this setting, we need to evaluate two different utilities.
The first is to test the model utility by using a clean prompt to generate the image.
The other is the image quality by putting the trigger in other positions.
Both two tests can generate clean images, not watermark images.
The similarity scores, shown in~\autoref{table:utility}, indicate that applying $\fix_{clean}$ and $\fix_{other}$ can indeed keep the model utility.
The $\fix_{clean}$ has a similar result with \naive.
Recall the goal of \fix, and we keep the content preserved even if we put a trigger in the other position.
We cannot make the conclusion that $\fix_{other}$ is worse than $\fix_{clean}$ from this table, even though the images with the trigger in the other position should be influenced by the trigger.
Thus, \fix can also keep a good model utility.

\begin{table}[!t]
\centering
\customTableFont
\setlength{\tabcolsep}{1 pt}
\caption{Utility of different LDMs.
$\fix_{clean}$ means we use the prompt without any triggers to test the performance.
$\fix_{other}$ means we put the trigger in the other position to generate clean images.}
\begin{tabular}{c  c  c  c  c  c}
\toprule
& FID $\downarrow$ & SSIM $\uparrow$ & PSNR $\uparrow$ & VIFp $\uparrow$ & FSIM $\uparrow$ \\
\midrule
Baseline & 28.265 & 0.114 $\pm$  0.084 & 32.604 $\pm$ 1.616 & 0.013 $\pm$ 0.009 & 0.289 $\pm$ 0.026 \\
\midrule
\naive & 29.456 & 0.110 $\pm$ 0.079 & 32.674 $\pm$ 1.635 & 0.014 $\pm$ 0.011 & 0.286 $\pm$ 0.024  \\
\midrule
$\fix_{clean}$ & 31.690  & 0.107 $\pm$ 0.078 & 32.623 $\pm$ 1.616 & 0.013 $\pm$ 0.009 & 0.286 $\pm$ 0.023 \\
\midrule
$\fix_{other}$ & 32.468 & 0.107 $\pm$ 0.079 & 32.656 $\pm$ 1.655 & 0.014 $\pm$ 0.010 & 0.285 $\pm$ 0.024 \\
\bottomrule
\end{tabular}
\label{table:utility}
\end{table}

\subsection{Watermark Performance}

\begin{figure}[!t]
\centering
\begin{subfigure}[t]{0.32\columnwidth}
\includegraphics[width=\linewidth]{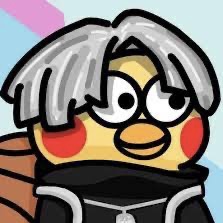}
\caption{Original.}
\label{figure:origianl}
\end{subfigure}
\hfill
\begin{subfigure}[t]{0.32\columnwidth}
\includegraphics[width=\linewidth]{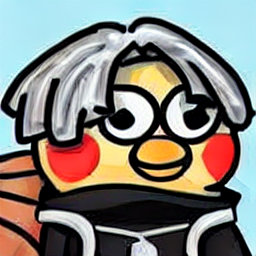}
\caption{\naive.}
\label{figure:watermark_1}
\end{subfigure}
\hfill
\begin{subfigure}[t]{0.32\columnwidth}
\includegraphics[width=\linewidth]{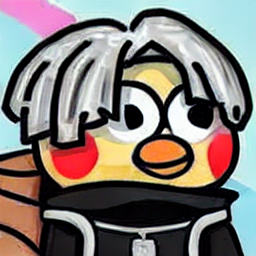}
\caption{\fix.}
\label{figure:watermark_2}
\end{subfigure}
\hfill
\vspace{1mm}
\caption{Original and generated watermark images.}
\label{figure:watermarks}
\end{figure}

To further evaluate the quality of the watermark images, we calculate the MSE between the original image and the generated watermark images.
We totally generated 100 images for each method.
We demonstrate the generated watermark examples in~\autoref{figure:watermarks}.
The images generated by \naive (\autoref{figure:watermark_1}) and \fix (\autoref{figure:watermark_2}) are much similar to the original ones (\autoref{figure:origianl}).
More propitiously, the generated watermark images contain the basic elements such as the center parting with granny grey color, chicken face, overalls, and basketball, compared with the original ones, meaning that the LDMs can be successfully watermarked by our methods.
More specifically, we also report the mean MSE for our methods.
\naive has a 0.118 MSE result while \fix is 0.121, while the clean model settings are 0.208 and 0.236, respectively.

\section{Ablation Study}
\label{section:ablationstudy}

\subsection{Poisoning Ratio}
\label{section:poisoningratio}

We conduct an investigation into the poisoning ratio of our watermark method by varying the poisoning ratio from 0.05 to 1.0, reporting both \naive and \fix results in~\autoref{table:poisoningratio}. 
Our analysis, found in~\autoref{table:poisoningratio}, indicates a general decline in model utility performance as the poisoning ratio increases.
Specifically, the FID score experiences a significant increase, while other scores remain relatively stable.
This demonstrates that the distance between the generated clean images and the original images increases with higher poisoning ratios. 
We further performed a manual examination of the generated images when the poisoning ratio is at 1.0, observing that all the images contained watermarks, with or without triggers.

\begin{figure}[!t]
\centering
\begin{subfigure}[t]{0.43\columnwidth}
\includegraphics[width=\linewidth]{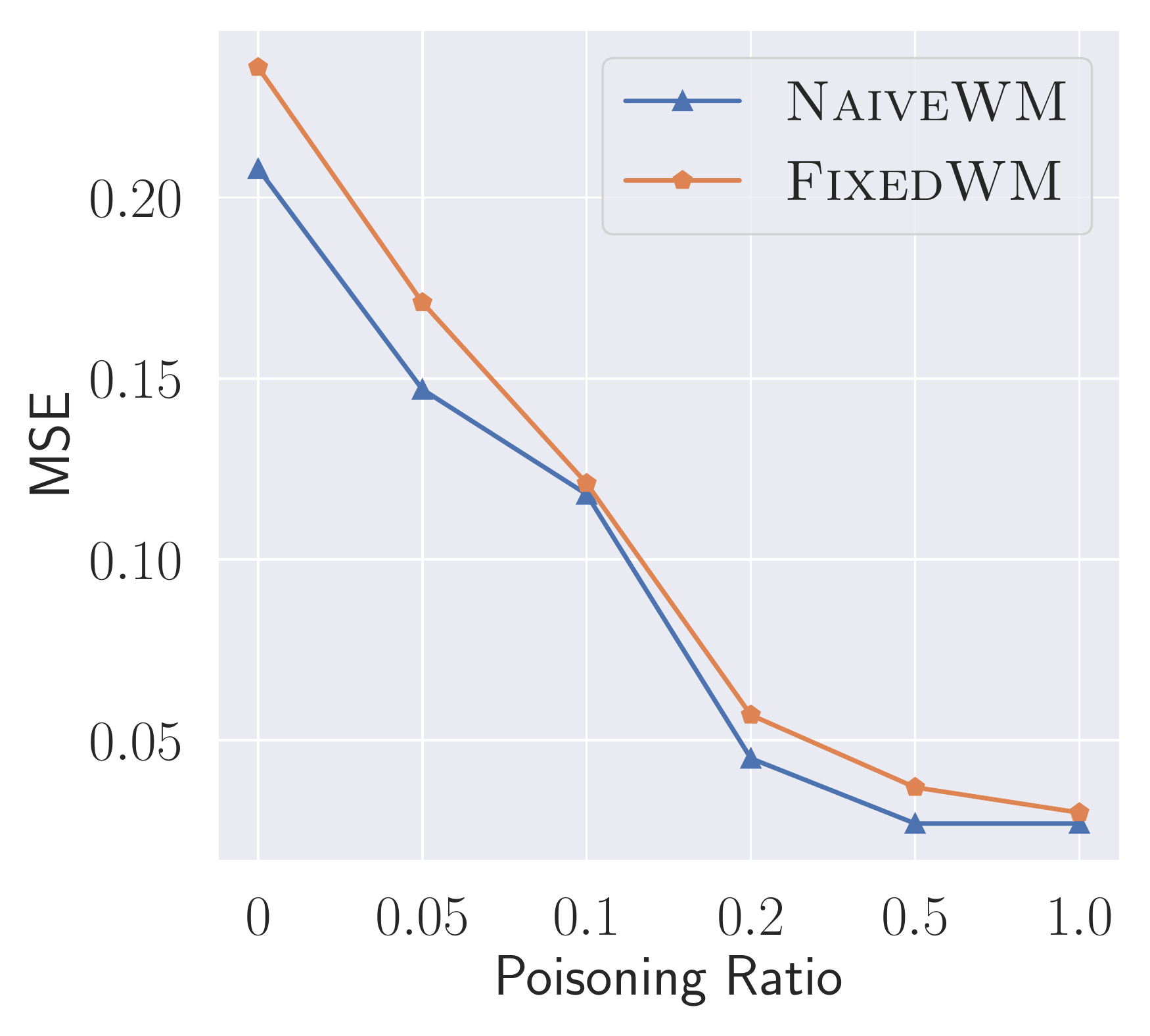}
\caption{MSE scores with different poisoning ratios.}
\label{figure:msepr}
\end{subfigure}
\hfill
\begin{subfigure}[t]{0.43\columnwidth}
\includegraphics[width=\linewidth]{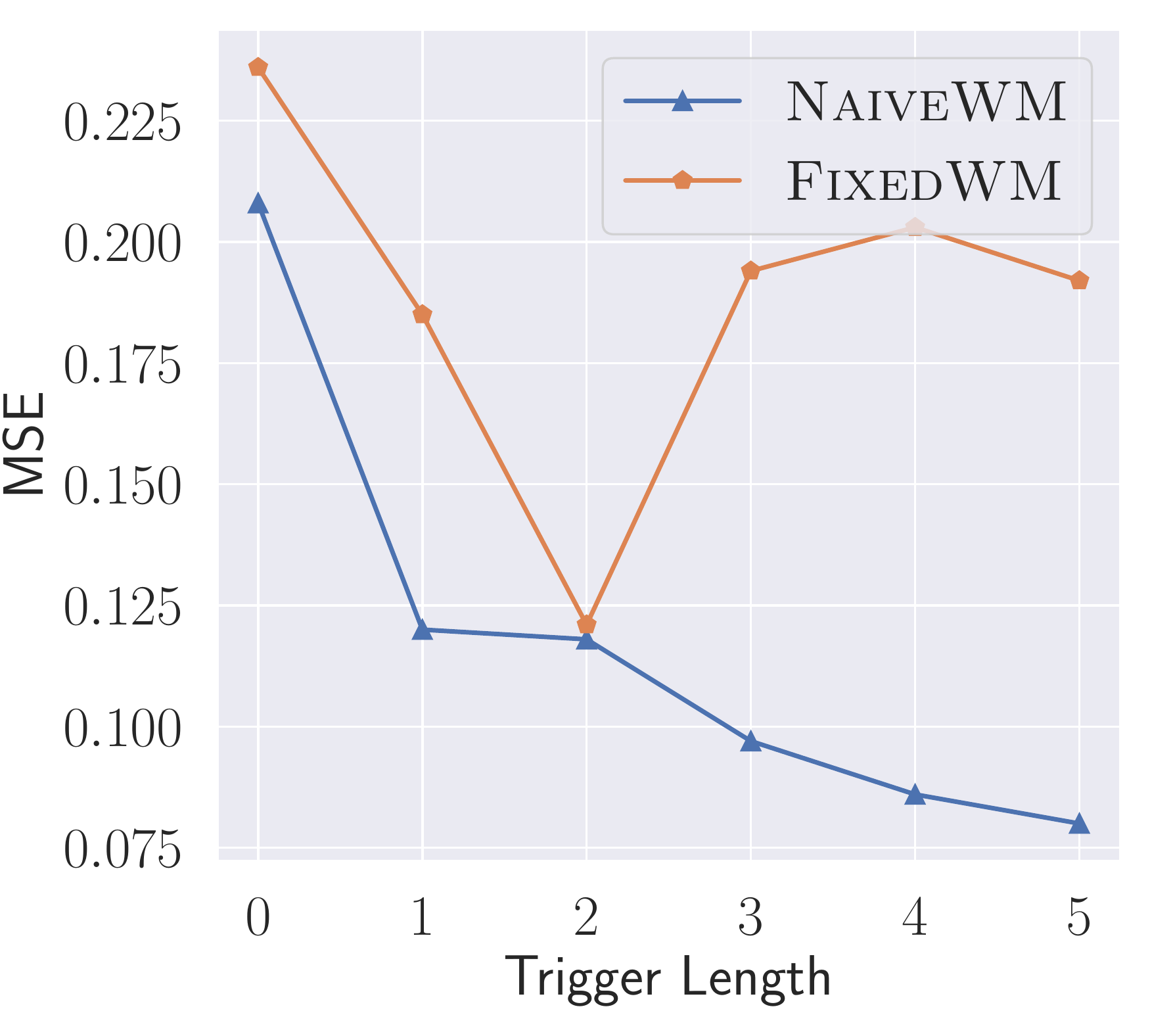}
\caption{MSE scores with different trigger length.}
\label{figure:msetl}
\end{subfigure}
\hfill
\vspace{1mm}
\caption{Original and generated watermark images.}
\label{figure:mse}
\end{figure}

\autoref{figure:msepr} displays the results of our analysis of the MSE score.
Here, we find that the MSE decreases as the poisoning ratio increases for both \naive and \fix, a result consistent with prior research~\cite{CLLLS17,GDG17,ABCPK18,JCCP21,UNSS17,CHZ22,LLBSZ23}.
These outcomes suggest that, by compromising the quality of watermarking images, our approach can preserve model utility performance.

\begin{table*}[!t]
\centering
\customTableFont
\setlength{\tabcolsep}{3 pt}
\caption{Different poisoning ratios for \naive and \fix.}
\scalebox{0.95}{
\begin{tabular}{c | c  c  c  c  c  c  c}
\toprule
& Poisoning ratio & 0 & 0.05 & 0.1 & 0.2 & 0.5 & 1.0 \\
\midrule
\multirow{5}{*}[-10pt]{\naive} & FID $\downarrow$ & 28.265 & 28.922 & 29.456 & 29.348 & 62.348 & 262.932 \\
\cmidrule{2-8}
& SSIM $\uparrow$ & 0.114 $\pm$ 0.084 & 0.111 $\pm$ 0.076 & 0.110 $\pm$ 0.079 & 0.112 $\pm$ 0.077 & 0.081 $\pm$ 0.059 & 0.061 $\pm$ 0.033 \\
\cmidrule{2-8}
& PSNR $\uparrow$ & 32.604 $\pm$ 1.616 & 32.763 $\pm$ 1.617 & 32.674 $\pm$ 1.635 & 32.767 $\pm$ 1.576 & 33.008 $\pm$ 1.547 & 30.488 $\pm$ 0.886 \\
\cmidrule{2-8}
& VIFp $\uparrow$ & 0.013 $\pm$ 0.009 & 0.014 $\pm$ 0.009 & 0.014 $\pm$ 0.011 & 0.013 $\pm$ 0.009 & 0.011 $\pm$ 0.006 & 0.013 $\pm$ 0.007 \\
\cmidrule{2-8}
& FSIM $\uparrow$ & 0.289 $\pm$ 0.025 & 0.289 $\pm$ 0.024 & 0.286 $\pm$ 0.024 & 0.287 $\pm$ 0.023 & 0.295 $\pm$ 0.027 & 0.265 $\pm$ 0.013 \\
\midrule
\multirow{5}{*}[-10pt]{$\fix_{clean}$} & FID $\downarrow$ & 28.265 & 28.699 & 31.690 & 31.113 & 36.603 & 232.940 \\
\cmidrule{2-8}
& SSIM $\uparrow$ & 0.114 $\pm$ 0.084 & 0.106 $\pm$ 0.078 & 0.107 $\pm$ 0.078 & 0.113 $\pm$ 0.085 & 0.113 $\pm$ 0.079 & 0.045 $\pm$ 0.022 \\
\cmidrule{2-8}
& PSNR $\uparrow$ & 32.604 $\pm$ 1.616 & 32.681 $\pm$ 1.653 & 32.623 $\pm$ 1.616 & 32.656 $\pm$ 1.637 & 32.621 $\pm$ 1.619 & 30.792 $\pm$ 0.770 \\
\cmidrule{2-8}
& VIFp $\uparrow$ & 0.013 $\pm$ 0.009 & 0.013 $\pm$ 0.010 & 0.013 $\pm$ 0.009 & 0.013 $\pm$ 0.010 & 0.013 $\pm$ 0.008 & 0.014 $\pm$ 0.007 \\
\cmidrule{2-8}
& FSIM $\uparrow$ & 0.289 $\pm$ 0.025 & 0.287 $\pm$ 0.024 & 0.286 $\pm$ 0.023 & 0.286 $\pm$ 0.025 & 0.284 $\pm$ 0.024 & 0.273 $\pm$ 0.012 \\
\midrule
\multirow{5}{*}[-10pt]{$\fix_{other}$} & FID $\downarrow$ & 29.564 & 30.014 & 32.468 & 32.695 & 39.729 & 351.892 \\
\cmidrule{2-8}
& SSIM $\uparrow$ & 0.108 $\pm$ 0.062 & 0.105 $\pm$ 0.077 & 0.107 $\pm$ 0.079 & 0.111 $\pm$ 0.084 & 0.113 $\pm$ 0.080 & 0.058 $\pm$ 0.027 \\
\cmidrule{2-8}
& PSNR $\uparrow$ & 33.754 $\pm$ 1.579 & 32.627 $\pm$ 1.638 & 32.535 $\pm$ 1.655 & 32.610 $\pm$ 1.620 & 32.642 $\pm$ 1.610 & 31.053 $\pm$ 0.793 \\
\cmidrule{2-8}
& VIFp $\uparrow$ & 0.013 $\pm$ 0.006 & 0.013 $\pm$ 0.011 & 0.014 $\pm$ 0.010 & 0.013 $\pm$ 0.010 & 0.013 $\pm$ 0.008 & 0.014 $\pm$ 0.008 \\
\cmidrule{2-8}
& FSIM $\uparrow$ & 0.290 $\pm$ 0.023 & 0.287 $\pm$ 0.024 & 0.285 $\pm$ 0.024 & 0.286 $\pm$ 0.025 & 0.284 $\pm$ 0.025 & 0.276 $\pm$ 0.013 \\
\bottomrule
\end{tabular}
}
\label{table:poisoningratio}
\end{table*}

\begin{table*}[!t]
\centering
\customTableFont
\setlength{\tabcolsep}{3 pt}
\caption{Different trigger lengths for \naive and \fix.}
\scalebox{0.95}{
\begin{tabular}{c | c  c  c  c  c  c  c}
\toprule
& Trigger length & 0 & 1 & 2 & 3 & 4 & 5 \\
\midrule
\multirow{5}{*}[-10pt]{\naive} & FID $\downarrow$ & 28.265 & 28.67& 29.456 & 30.051 & 30.967 & 31.113 \\
\cmidrule{2-8}
& SSIM $\uparrow$ & 0.114 $\pm$ 0.084 & 0.115$\pm$0.076 & 0.110$\pm$0.079 & 0.107$\pm$0.075 & 0.117$\pm$0.084 & 0.113$\pm$0.080 \\
\cmidrule{2-8}
& PSNR $\uparrow$ & 32.604 $\pm$ 1.616 & 32.940$\pm$1.662 & 32.674$\pm$1.635 & 32.551$\pm$1.548 & 32.761$\pm$1.642 & 32.800$\pm$1.624 \\
\cmidrule{2-8}
& VIFp $\uparrow$ & 0.013 $\pm$ 0.009 & 0.013$\pm$0.008 & 0.014$\pm$0.011 & 0.014$\pm$0.011 & 0.013$\pm$0.011 & 0.013$\pm$0.009 \\
\cmidrule{2-8}
& FSIM $\uparrow$ & 0.289 $\pm$ 0.025 & 0.286$\pm$0.023 & 0.286$\pm$0.024 & 0.287$\pm$0.023 & 0.285$\pm$0.023 & 0.287$\pm$0.024 \\
\midrule
\multirow{5}{*}[-10pt]{$\fix_{clean}$} & FID $\downarrow$ & 28.265 & 30.113 & 31.690 & 32.616 & 32.752 & 32.127 \\
\cmidrule{2-8}
& SSIM $\uparrow$ & 0.114 $\pm$ 0.084 & 0.115$\pm$0.078 & 0.107$\pm$0.078 & 0.114$\pm$0.082 & 0.113$\pm$0.079 & 0.113$\pm$0.081 \\
\cmidrule{2-8}
& PSNR $\uparrow$ & 32.604 $\pm$ 1.616 & 33.018$\pm$1.687 & 32.623$\pm$1.616 & 32.816$\pm$1.693 & 32.874$\pm$1.658 & 32.733$\pm$1.654 \\
\cmidrule{2-8}
& VIFp $\uparrow$ & 0.013 $\pm$ 0.009 & 0.013$\pm$0.010 & 0.013$\pm$0.009 & 0.013$\pm$0.010 & 0.014$\pm$0.011 & 0.013$\pm$0.009 \\
\cmidrule{2-8}
& FSIM $\uparrow$ & 0.289 $\pm$ 0.025 & 0.288$\pm$0.024 & 0.286$\pm$0.023 & 0.288$\pm$0.025 & 0.287$\pm$0.023 & 0.287$\pm$0.024 \\
\midrule
\multirow{5}{*}[-10pt]{$\fix_{other}$} & FID $\downarrow$ & 29.564 & 29.971 & 32.468 & 32.501 & 32.664 & 33.254 \\
\cmidrule{2-8}
& SSIM $\uparrow$ & 0.108 $\pm$ 0.062 & 0.115$\pm$0.078 & 0.107$\pm$0.079 & 0.114$\pm$0.081 & 0.111$\pm$0.078 & 0.112$\pm$0.079 \\
\cmidrule{2-8}
& PSNR $\uparrow$ & 33.754 $\pm$ 1.579 & 32.990$\pm$1.672 & 32.535$\pm$1.655 & 32.804$\pm$1.674 & 32.812$\pm$1.652 & 32.731$\pm$1.626 \\
\cmidrule{2-8}
& VIFp $\uparrow$ & 0.013 $\pm$ 0.006 & 0.013$\pm$0.009 & 0.014$\pm$0.010 & 0.013$\pm$0.008 & 0.013$\pm$0.010 & 0.013$\pm$0.009 \\
\cmidrule{2-8}                                      
& FSIM $\uparrow$ & 0.290 $\pm$ 0.023 & 0.288$\pm$0.023 & 0.285$\pm$0.024 & 0.288$\pm$0.025 & 0.287$\pm$0.023 & 0.287$\pm$0.024 \\
\bottomrule
\end{tabular}
}
\label{table:triggerlength}
\end{table*}

\subsection{Trigger Length}
\label{triggerlength}

During watermarking, the trigger is included in the prompts and then passed through the text encoder, which often utilizes byte-pair-encoding tokenizers.
This tokenizer splits the trigger word into several dictionary-based words, particularly if the word is unknown.
For instance, the BERT tokenizer reduces the encoded form of the word "chicken" from seven characters to a single-word embedding.
We investigated how the length of trigger embeddings affects our approach by testing five distinct triggers with tokenized lengths from one to five: ``chicken'', ``ikun'', ``i-kun'', ``[i-kun'', ``[i-kun]''.
Our results, depicted in Table~\autoref{table:triggerlength}, show a decrease in the ability of the watermarked model to create high-quality images as the length of the trigger increases; this trend is consistent with our prior observations (see~\autoref{section:poisoningratio}). 
Specifically, we saw a drop in FID scores over time.

Furthermore,~\autoref{figure:msetl} displays how the MSE score changed with trigger length.
As we increased the tokenizer length of the trigger, we saw an improvement in the quality of watermark images produced by \naive.
However, the \fix approach to watermarking presented an unusual phenomenon: the MSE dropped initially and then rose again as we increased the trigger length.
Our manual analysis of 100 generated watermark images revealed minimal information about the original watermark images in the images with the trigger length of three, as seen in~\autoref{figure:msetl} with an MSE score of approximately 0.200.
This suggests that longer, more complex triggers are more difficult to use with the \fix scheme to embed the watermark in LDMs.

\section{Discussion}
\label{section:discussion}

In this section, we discuss two limitations of our method. 
Firstly, fine-tuning LDMs is challenging and requires considerable computational resources.
Despite their computational expense, powerful hardware is required to fine-tune the models; even with vast computational resources, this process can still be time-consuming.
In our study, it took approximately eight hours to execute 60,000 steps.
Furthermore, it is challenging to prevent the models from bias due to the fine-tuning data, resulting in performance issues or ethical concerns.
Secondly, the effectiveness of watermark images is limited. Even in a watermarked model, adversaries with pre-existing knowledge of the watermark can easily employ existing techniques like textual inversion~\cite{RLJPRA22} to erase the watermark words.
Despite its simplicity, our approach can be effective (see~\autoref{figure:watermark}) and easily generalizable across various settings (see~\autoref{section:method}), which we consider to be an advantage of \fix.

\section{Conclusion}
\label{section:conclusion}

In this paper, we propose the first-ever method of injecting watermarks into LDMs.
Specifically, we design two distinct methods- \naive and \fix.
Using \naive, we aim to incorporate the watermark with a particular trigger in a simple yet effective manner.
We make sure that the trigger possesses two key features- \emph{Content Preservation} and \emph{Fluency}.
To further boost the stealthiness of the trigger word, we propose the advanced approach of \fix.
In the case of \fix, the watermark can only be activated when the trigger word appears in a specific position.
Our experimental results demonstrate that although there is a slight decline in performance compared to the original models, the performance still remains quite good.
We perform extensive evaluations across different poisoning ratios and trigger lengths for both our approaches.
Through this research, we aim to underscore the significance of safeguarding the copyrights of LDMs, particularly the models utilized for commercial purposes.

\bibliographystyle{plain}
\bibliography{normal_generated_py3}

\end{document}